\documentclass[11pt]{article}

\usepackage{amsmath}
\usepackage{amssymb}
\usepackage{cancel}

\numberwithin{equation}{section}

\begin{document}
\baselineskip=18pt

\begin{titlepage}
\begin{flushright}
KYUSHU-HET-122\\[-1mm]
KUNS-2233
\end{flushright}

\begin{center}
\vspace*{20mm}

{\Large\bf 
Geometry-free neutrino masses in curved spacetime
}
\vspace*{12mm}

Atsushi Watanabe$^*$,
%\footnote{watanabe@higgs.phys.kyushu-u.ac.jp}
~Koichi Yoshioka$^\dag$
%\footnote{yoshioka@gauge.scphys.kyoto-u.ac.jp}
\vspace*{3mm}

$^*${\small {\it 
Department of Physics, Kyushu University, Fukuoka 812-8581, Japan}}\\
$^\dag${\small {\it Department of Physics, Kyoto University, Kyoto 
606-8502, Japan}}\\

\vspace*{5mm}

{\small (October, 2009)}
\end{center}

\vspace*{5mm}

\begin{abstract}\noindent%
The seesaw-induced neutrino mass is discussed 
in a generic class of curved spacetime, including the flat and warped
extra dimensions. For Majorana masses in the bulk and on the boundary,
the exact forms of seesaw-induced masses are derived by using the
Kaluza-Klein mode expansion and the lepton number violating correlator
for bulk fermion. It is found that the neutrino mass is determined
without the knowledge of wave functions and whole background geometry
when the metric factor is fixed on the boundary, e.g.\ by solving the
hierarchy problem.
\end{abstract} 

\end{titlepage}
\newpage
%%%%%%%%%%%%%%%%%%%%%%%%%%%%%%%%%%%%%%%%%%%%%%%%%%%%%%%%%%%%%%%%%%%%%%%
\section{Introduction}
%%%%%%%%%%%%%%%%%%%%%%%%%%%%%%%%%%%%%%%%%%%%%%%%%%%%%%%%%%%%%%%%%%%%%%%

Over the past decade, theories with extra dimensions have attracted
great attention as feasible paradigm to understand the unresolved
problems in the Standard Model (SM)\@. For example, the
four-dimensional Planck scale becomes effective, which is made out of
the fundamental scale of higher-dimensional gravity and the large
volume size of extra space by which the matter-gravity 
coupling is weakened~\cite{ADD}. The localized gravity with the warped 
metric~\cite{RS} also provides a framework for solving the gauge
hierarchy by small overlap between matter and gravitational fields.

An interesting aspect of higher-dimensional theory is the interplay
with the neutrino physics. As in the same way that the flux is diluted
in the extra space, the low-energy neutrino mass receives a volume
suppression if right-handed neutrinos live in the extra
dimensions~\cite{Psi}. It has also been discussed that the
localization of bulk fermions in the warped extra dimension produces
tiny neutrino masses~\cite{bulkF}. Such a connection between the
neutrino physics and extra dimensions is a subject of great interests
to particle physics.

In this letter, we explore the bulk Majorana neutrinos in a generic
type of non-factorizable geometry, including the flat and warped extra
dimensions. According to the location of lepton-number violation, we
separately consider two types of Majorana mass terms consistent with
the Lorentz invariance: (i) in the bulk and (ii) on the Planck
brane. (The TeV-brane mass term is a straightforward 
application.) \ While the exact form of mass spectrum and wave
functions cannot be found for bulk fermions, the low-energy effective
mass for left-handed neutrinos is obtained analytically.

\bigskip

%%%%%%%%%%%%%%%%%%%%%%%%%%%%%%%%%%%%%%%%%%%%%%%%%%%%%%%%%%%%%%%%%%%%%%%
\section{Majorana mass on the Planck brane}
%%%%%%%%%%%%%%%%%%%%%%%%%%%%%%%%%%%%%%%%%%%%%%%%%%%%%%%%%%%%%%%%%%%%%%%

Throughout this letter, we consider the five-dimensional theory on the 
gravitational background with the following non-factorizable form
\begin{equation}
ds^2 \;=\; g_{MN} dx^M dx^N \;=\; 
\rho^{-2}(y)\,\eta_{ab}dx^a dx^b -dy^2,
\label{metric}
\end{equation}
where $\eta_{ab}$ is the four-dimensional Minkowski metric. The
fifth dimension has two orbifold fixed points 
at $y=0$ and $y=L$\@. A phenomenologically interesting example is the
anti de-Sitter (AdS) space where the metric factor $\rho(y)$ is given 
by $e^{k|y|}$ ($k$ is the AdS curvature). In this case, the
physical scales for $y=0$ and $y=L$ boundaries are differently
set to the Planck and TeV scales with the warp factor $e^{-kL}$, and
they are referred to as the Planck and TeV branes. An index 
of $\mathcal{O}(10)$, namely $kL\simeq 37$, is sufficient for
obtaining the electroweak/Planck mass hierarchy.

The SM fields, especially the left-handed neutrinos $N$ and the
Higgs field $H$ are assumed to be localized at $y=L$. The right-handed
neutrinos are introduced as bulk Dirac fermions $\Psi(x,y)$ which obey
the boundary 
conditions $\Psi(x,-y)=\gamma_5\Psi(x,y)$ and $\Psi(x,L-y)=\gamma_5
\Psi(x,L+y)$. Let us consider the Majorana mass on the Planck
brane. The Planck-brane Majorana mass on the warped 
AdS$_5$ background has been discussed in various
setups~\cite{braneM}. In the general non-factorizable
geometry~\eqref{metric}, the Lagrangian relevant for neutrino physics
is given by
\begin{eqnarray}
\mathcal{L} &=& 
\sqrt{g} \Big[\, i\overline{\Psi}\Gamma^M D_M\Psi
- m_d \theta(y) \overline{\Psi}\Psi
- \Big(\frac{1}{2}M \overline{\Psi^c}\Psi
+ {\rm h.c.}\Big)\delta(y) 
\nonumber\\
&&\qquad\qquad\qquad
+\left[ \, i \overline{N} \gamma^\mu\partial_\mu N 
- \rho\,\big(m \overline{\Psi} N 
+ {\rm h.c.}\big) \right]
\delta(y-L) \, \Big],
\label{Lag1}
\end{eqnarray}
where $D_M$ is the covariant derivative which includes the spin
connection, and the gamma matrices are related as
$\Gamma^\mu=\gamma^\mu$ and $\Gamma^y=i\gamma_5$. The step 
function $\theta(y)$ is needed since the mass 
operator $\overline{\Psi}\Psi$ is odd under the reflection parities
with respect to $y=0$ and $y=L$. The charge-conjugated 
spinor $\Psi^c$ is defined 
as $\Psi^c=\Gamma^3\Gamma^1 \overline{\Psi}^{\rm T}$ such that
it becomes Lorentz covariant in five dimensions. The 
parameter $m$ means the electroweak-breaking Dirac mass given by the
vacuum expectation value $\langle H \rangle$. In the above Lagrangian,
we have rescaled the Higgs field so that its kinetic term becomes
canonical. That leads to the $\rho$ factor in the boundary mass term,
and hence $m$ is regarded as a parameter of the electroweak scale.

The low-energy effective theory for neutrino masses is deduced by
usual Kaluza-Klein (KK) expansion. After rescaling $\Psi$ and $N$ so
that their kinetic terms become canonical, we expand the right-handed
neutrino fields as
\begin{equation}
\Psi(x,y) \;=\; \left(\begin{array}{c}
\sum_n\chi_R^n(y) \psi_R^n(x) \\[2mm]
\sum_n\chi_L^n(y) \psi_L^n(x)
\end{array}\right),
\end{equation}
with the wave functions $\chi_{R,L}^n(y)$ which satisfy the equations
of motion
\begin{eqnarray}
\Big( \partial_y + m_d\theta(y) 
- \frac{\partial_y\rho}{2\rho} \Big)
\chi_R^n &=& +M_{K_n}\rho(y)\chi_L^n, 
\label{eq1}\\
\Big( \partial_y - m_d\theta(y) 
- \frac{\partial_y\rho}{2\rho} \Big)
\chi_L^n &=& -M_{K_n}\rho(y)\chi_R^n, 
\label{eq2}
\end{eqnarray}
where $M_{K_n}$ represent the KK mass eigenvalues. The non-trivial
solution for the zero mode is given by
\begin{eqnarray}
  \chi_R^0(y) \;=\; A\,e^{-m_d|y|}\rho^{\frac{1}{2}}(y)\,.
\end{eqnarray}
The normalization constant $A$ is fixed by the 
condition $\int_0^L\!dy\,(\chi_R^0)^2=1$. The localization profile of
the zero mode is fixed by the bulk Dirac mass $m_d$ and the metric
factor.

Substituting the KK-mode expansion into~\eqref{Lag1} and
integrating it over the extra space, we obtain the four-dimensional
effective masses for the normalized Majorana fermions:
\begin{eqnarray}
\mathcal{M}
=
\left( \begin{array}{c|c}
 0 &  \;\; \mathcal{M}_D^{\rm T} \,\, \\\hline
 &  \\
\!\!\mathcal{M}_D\! &  \;\; \mathcal{M}_H  \,\, \\
 &  \\
 &  \\
\end{array}
\right)
=  
\left( \begin{array}{c|cccc}
 & m_0^{\rm T} & m_1^{\rm T} &  & \cdots \\\hline
m_0 & -M_{R_{00}}^* & -M_{R_{01}}^* &  & \cdots \\
m_1 & -M_{R_{10}}^* & -M_{R_{11}}^* & M_{K_1} & \cdots \\
 &  & \phantom{-}M_{K_1} &  & \cdots \\
\vdots & \vdots & \vdots & \vdots & \ddots \\
\end{array}
\right)\!, \quad
\mathcal{N}
=
\begin{pmatrix}
\nu_L^{} \\[.5mm]  \epsilon\psi_R^0{}^* \\[.5mm]
\epsilon \psi_R^1{}^* \\[.5mm]  \psi_L^1 \\
\vdots \\
\end{pmatrix}\!,
\label{KKmm}
\end{eqnarray}
\begin{eqnarray}
M_{R_{nm}} \,=\, \rho^{-1}(0)\,
\chi_R^n{}^{\rm T}(0) M \chi_R^m(0),
\quad\quad
m_n \,=\, \chi_R^n{}^\dag(L)\,m,
\end{eqnarray}
where $\nu_L^{}$ is the left-handed neutrinos in the two-component
notation: $N(x)=
\big(\begin{smallmatrix}0 \\ \nu_L^{}\end{smallmatrix}\big)$. Besides
the upper-left entry, the vanishing elements in the mass 
matrix $\mathcal{M}$ come from the Dirichlet 
conditions $\chi_L^n(0)=\chi_L^n(L)=0$.

The low-energy mass spectrum is obtained by the diagonalization 
of $\mathcal{M}$. Under the assumption $m_n \ll M_{R_{nm}}$, the
Majorana mass matrix for the left-handed neutrinos is
approximated as $M_\nu=-\mathcal{M}_D^{\rm T} \mathcal{M}_H^{-1}
\mathcal{M}_D^{}$. The inverse of the heavy-sector infinite
matrix $\mathcal{M}_H$ is found
\begin{eqnarray}
\mathcal{M}_H^{-1} 
\;=\; 
\left( \begin{array}{cccccc}
\frac{-1}{M_{R_{00}}^*} & & 
\!\!\!\frac{M_{R_{01}}^*}{M_{R_{00}}^*}\frac{-1}{M_{K_1}}\!\!\!
& & 
\!\!\!\frac{M_{R_{02}}^*}{M_{R_{00}}^*}\frac{-1}{M_{K_2}}\!\!
& \cdots \\
& & \frac{1}{M_{K_1}} & & & \cdots \\
\!\!\frac{M_{R_{01}}^*}{M_{R_{00}}^*}\frac{-1}{M_{K_1}} & 
\frac{1}{M_{K_1}} & & & & \cdots \\
& & & & \frac{1}{M_{K_2}} & \cdots \\
\!\!\frac{M_{R_{02}}^*}{M_{R_{00}}^*}\frac{-1}{M_{K_2}} & & &
\frac{1}{M_{K_2}} & & \cdots \\
\vdots & \vdots & \vdots & \vdots & \vdots & \ddots \\
\end{array}
\right).
\label{inv}
\end{eqnarray}
Multiplying ${\cal M}_D$ from both sides, we find that all the KK-mode
contributions vanish, except for the zero-mode one:
\begin{eqnarray}
M_\nu \;=\; \frac{m_0^{\rm T}m_0}{M_{R_{00}}^*} \;=\;
e^{-2m_dL}\rho(L)\,
\frac{m^{\rm T}m}{M^*}.
\label{mnu}
\end{eqnarray}
The seesaw-induced mass has been studied by
truncating the large matrix $\mathcal{M}$ to include finite numbers of
KK modes. However the infinite-dimensional matrix~\eqref{inv} shows
that the exact form of KK-induced Majorana masses is obtained without
the knowledge of mass spectrum and wave functions of KK-excited modes,
which are generally complicated to perform the seesaw operation.

The effective mass $M_\nu$ involves only the zero-mode piece and the
other ones are cancelled out. This is because the lepton number is
violated only in the zero-mode sector, which is clearly seen in the
matrix~\eqref{inv} whose elements conserve the lepton number except
for the zero-mode one, $-1/M_{R_{00}}^*$. One can also see this by
making the mixed states between $\psi_R^0$ and $\psi_R^n$ with the
rotation angles $\tan\theta_n=\chi_R^n(0)/\chi_R^0(0)$. With this
change of basis, the lepton-number violating terms in the heavy-sector 
matrix $\mathcal{M}_H$ are all erased out besides the zero-mode 
entry. In the case of localized Majorana masses,
therefore, only the zero mode $\psi_R^0$ takes part in the seesaw
mechanism. It should be noted that $\psi_R^0$ is not a mass eigenstate
in four-dimensional viewpoint. The real ``zero mode'', which would be
observed in future particle experiments, is a linear combination of KK
modes.

The suppression of neutrino mass~\eqref{mnu} to the eV range is
achieved with appropriate bulk masses and gravitational
background. The zero modes of bulk neutrinos should be localized
towards the TeV brane, which reduces the Planck-brane Majorana mass to
an intermediate seesaw scale. If the metric factor $\rho(L)$ is used
to resolve the gauge hierarchy, $M\rho^{-1}(L)$ is around or 
smaller than the TeV scale.\footnote{This conclusion can be changed by
considering a bulk Higgs field and/or a small Higgs mass, which is
stabilized by other dynamics such as supersymmetry.} \ Therefore the
wave function factor with a (positive) non-vanishing $m_d$ is suitable
for having tiny neutrino masses.
To be more concrete, let us consider the warped AdS$_5$ metric
($\rho(y)=e^{k|y|}$) and parameterize the bulk Dirac and Majorana
masses as $m_d=c_d k$ and $M=c_M k$. From~\eqref{mnu}, it follows
that a coefficient $c_d\simeq 0.38$ 
produces $\mathcal{O}(0.1)\;{\rm eV}$ neutrino mass 
for $c_M=1$. Around a suppressed value $c_M\sim 10^{-4}$, the
coefficient $c_d$ reaches the de-localized limit $c_d=0.5$ and the zero
mode starts to be localized towards the Planck brane.

Another way to realize a tiny Majorana neutrino mass is to suppress the
boundary Dirac mass (i.e.\ the Yukawa coupling). That is related to the
charged-lepton sector on the TeV brane. One may apply some
mechanisms which have been proposed in four or higher-dimensional
theory to solve the fermion mass problem, e.g.\ with an abelian flavor
symmetry $\grave{a}~la$ Froggatt and Nielsen~\cite{FN}. A different
approach is to assume that the left-handed neutrinos (lepton
doublets) and/or the right-handed charged leptons also access the
extra dimension. For an example where the right-handed charged leptons
propagate in the warped extra dimension, the bulk Dirac 
masses $(m_{d_e},m_{d_\mu},m_{d_\tau})=
(0.842,0.688,0.603)k$ reproduce the observed charged-lepton masses
for unit Yukawa couplings. If the lepton doublets also feel the extra
dimension, the zero modes of right-handed neutrinos should get closer
to the TeV brane for compensating the suppression by the left-handed
neutrinos.

\bigskip

%%%%%%%%%%%%%%%%%%%%%%%%%%%%%%%%%%%%%%%%%%%%%%%%%%%%%%%%%%%%%%%%%%%%%%%
\section{Majorana mass in the bulk}
%%%%%%%%%%%%%%%%%%%%%%%%%%%%%%%%%%%%%%%%%%%%%%%%%%%%%%%%%%%%%%%%%%%%%%%

Another interesting possibility for neutrino physics, though less
considered, is to put the Majorana mass in the five-dimensional bulk.
As in the previous section, the left-handed neutrinos $N$ and the
Higgs field $H$ are assumed to reside on the TeV brane. In addition,
we introduce the five-dimensional right-handed neutrinos $\Psi$ with
the Majorana mass term in the bulk. The Lagrangian on the general
non-factorizable background~\eqref{metric} is given by
\begin{eqnarray}
\mathcal{L} &=& 
\sqrt{g} \Big[\, 
 i\overline{\Psi}\Gamma^M D_M\Psi
- m_d \theta(y) \overline{\Psi}\Psi
- \Big(\frac{1}{2}M\overline{\Psi^c}\Psi
+ {\rm h.c.}\Big)
\nonumber\\
&&\qquad\qquad\qquad
+\left[ \, i\overline{N} \gamma^\mu\partial_\mu N 
- \rho\,\big(m \overline{\Psi} N 
+ {\rm h.c.}\big) \right]
\delta(y-L) \, \Big].
\end{eqnarray}
The bulk fields $\Psi$ are supposed to obey the boundary 
conditions $\Psi(x,-y)=\gamma_5\Psi(x,y)$ and $\Psi(x,L-y)=
\gamma_5\Psi(x,L+y)$ as in the previous section. Other types of
boundary conditions will be discussed later. The bulk Majorana mass in
the warped AdS$_5$ geometry has been mentioned with a different type
of compactification~\cite{bulkM}. In this letter, we do not consider
the Lorentz-violating Majorana mass of the 
form $\overline{\Psi^c}\gamma_5\Psi$, while often studied in the
literature.

For the bulk Majorana mass, it is not easy to find an analytic form
of the seesaw-induced mass by the KK expansion method. With the
eigenfunctions obeying~\eqref{eq1} and~\eqref{eq2}, the neutrino
mass matrix $\mathcal{M}$, which corresponds to~\eqref{KKmm}, becomes
\begin{eqnarray}
\mathcal{M} 
\,=\,
\left( \begin{array}{c|c}
 0 &  \;\; \mathcal{M}_D^{\rm T} \,\, \\\hline
 &  \\
\!\!\mathcal{M}_D\! &  \;\; \mathcal{M}_H  \,\, \\
 &  \\
\end{array}
\right)
\,=\,
\left( \begin{array}{c|cccc}
 & m_0^{\rm T} & m_1^{\rm T} &  & \cdots \\\hline
m_0 & -M_{R_{00}}^*  & -M_{R_{01}}^* & M_{L_{01}} & \cdots \\
m_1 & -M_{R_{10}}^*  & -M_{R_{11}}^* & M_{K_1} & \cdots \\
 & \phantom{-}M_{L_{10}} &  M_{K_1} & M_{L_{11}} & \cdots \\
\vdots & \vdots & \vdots & \vdots & \ddots \\
\end{array}
\right).
\end{eqnarray}
The KK-mode Majorana masses are determined by
\begin{equation}
M_{R_{nm}} \,=\, \int_0^L\!\!\! dy \,
\rho^{-1}\chi_R^n(y){}^{\rm T} M \chi_R^m(y),
\qquad
M_{L_{nm}} \,=\, \int_0^L\!\!\! dy \,
\rho^{-1} \chi_L^n(y){}^{\rm T} M \chi_L^m(y).
\end{equation}
The matrix ${\cal M}$ is too complicated for the seesaw integration to
be performed: the heavy-sector matrix ${\cal M}_H$ cannot be
obtained explicitly due to intricate wave functions. Moreover,
unlike the previous case, there is no trivially-vanishing element 
in $\mathcal{M}_H$ and it seems hard to write down its 
inverse $\mathcal{M}_H^{-1}$ and to evaluate the seesaw formula.

As an alternative to the KK-mode expansion, the propagator method for
bulk fermion is suitable for calculating low-energy neutrino
masses. For the present purpose, it is convenient to make a
non-canonical rescaling $\Psi\to\rho^2\Psi$ and the
Lagrangian is then rewritten as
\begin{eqnarray}
\mathcal{L}
\;=\;
\frac{1}{2}\overline{\Phi}D \Phi
\,+\sqrt{g}\left[\, i\overline{N}\gamma^\mu \partial_\mu N 
- \rho^3 \big(m\overline{\Psi} N 
+  {\rm h.c.}\big)\, \right]\delta(y-L),
\end{eqnarray}
where 
\begin{eqnarray}
D \,=\,
\begin{pmatrix}
i\rho\,\cancel{\partial} - \gamma_5 \partial_y 
- m_d\theta(y) & -M^* \\
-M & i\rho\,\cancel{\partial} - \gamma_5 \partial_y 
+ m_d\theta(y) \\
\end{pmatrix},
\qquad
\Phi \,=\,
\begin{pmatrix}
\Psi \\
\Psi^c \\
\end{pmatrix}.
\label{D}
\end{eqnarray}
The lepton number violating part of the two-point function is
extracted from the inverse of $D$. Regarding the
TeV-brane mass as a perturbation, we find the
effective Majorana mass for the canonically-normalized left-handed
neutrinos,
\begin{eqnarray}
M_\nu &=& \rho(L)\,m^{\rm T}
\langle \epsilon \nu_R^*(p,L)\nu_R^\dag(p,L) \rangle\big|_{p=0}\,m,
\label{pp}
\end{eqnarray}
where $\nu_R^{}$ is the right-handed component of (rescaled) bulk
spinor $\Psi$, 
and $p$ is the momentum in the four-dimensional Minkowski
spacetime. If one substitutes the KK-mode
expansion $\nu_R^{}=\sum_n\rho^\frac{-1}{2}\chi_R^n(y)\psi_R^n(p)$ into
the formula ($\rho^\frac{-1}{2}$ implies the above rescaling and
the canonical normalization of $\psi_R^n$),
\begin{eqnarray}
\rho(L)\,m^{\rm T} \langle\epsilon\nu_R^*(p,L)\nu_R^\dag(p,L)\rangle m
&=& \sum_{k,n}  m^{\rm T}\chi_R^k(L)^*
\langle \epsilon \psi_R^k(p)^* \psi_R^n(p)^\dag \rangle
\chi_R^n(L)^\dag m
\nonumber\\
&=& \sum_{k,n} m_k^{\rm T}
\bigg(\frac{{\cal M}_H^{\,*}}{p^2
-{\cal M}_H^{\,*}{\cal M}_H^{}}\bigg)_{\!\!R_{kn}}\!\! m_n\,,
\end{eqnarray}
where $R_{kn}$ means the matrix element for the KK right-handed
neutrinos $\psi_R^{k,n}$. The last line is the summation over all
heavy-mode contributions, just as performed in the previous
section. With the propagator at hand, therefore, one needs neither to
integrate over the extra space nor to diagonalize the infinite neutrino
matrix.

It is generally difficult to find the bulk Majorana
propagator. However in view of the seesaw scheme, we only need the
low-energy limit $p\to0$.\footnote{The zero-momentum limit is 
not actually required, but $\rho(y)p$ should be smaller than the 
fundamental scale at any point in the bulk for the following
procedure being valid.} \ It is seen from~\eqref{D} that, in the
low-energy limit, the metric factor $\rho$ vanishes away from the
problem and the propagator is found to have the same form as in the
flat extra dimension. Notice that the low-energy limit $p\to0$ is
allowed if the solution is not singular in this limit. For neutrino
physics, the Majorana mass lifts the chiral zero mode of bulk fermion
up to the heavy sector, and the propagator does not have a pole 
at $p=0$.

By solving the inverse of $D$ without $\cancel{\partial}$ parts
and setting the boundary conditions $\Psi(x,-y)=\gamma_5\Psi(x,y)$ and
$\Psi(x,L-y)=\gamma_5\Psi(x,L+y)$, we find the lepton number violating
correlator
\begin{eqnarray}
\langle \epsilon \nu_R^*(0,y)\nu_R^\dag(0,y') \rangle
&=& \frac{M}{(q^2-m_d^2)q\sinh(qL)}
\Big[ q\cosh(qy_<) - m_d \sinh(qy_<) \Big] \nonumber\\[1mm]
&& \qquad\qquad
\times \Big[ q\cosh(qy_>-qL) - m_d \sinh(qy_>-qL) \Big],
\label{propZp}
\end{eqnarray}
where $q^2=m_d^2 + |M|^2$ and
$y_<$ ($y_>$) stands for the lesser (greater) of $y$ and $y'$. From
the formula~\eqref{pp}, the Majorana mass is induced for 
left-handed neutrinos in the general non-factorizable geometry:
\begin{eqnarray}
M_\nu \;=\; \rho(L)\, 
\frac{ q \cosh(qL) - m_d \sinh(qL) }{\sinh(qL)}\;
\frac{m^{\rm T}m}{M^*}.
\label{mnuw1}
\end{eqnarray}
The typical heavy-mode scale is played by the bulk Majorana 
mass $M$ or the KK scale. Let us take $m_d=0$ for simplicity. For a
smaller value of Majorana mass, $qL\ll 1$, the neutrino mass is
approximated as $M_\nu \simeq \frac{\rho(L)}{L}\frac{m^2}{M^*}$. In
the KK-mode picture, the contribution from low-lying states, which is
governed by the effective Majorana mass $M\rho(L)^{-1}$, dominates
the seesaw-induced mass. In the opposite limit $qL\gg 1$, the neutrino
mass follows $M_\nu \simeq\frac{\rho(L)}{L}
\frac{m^2}{(1/L)}$, and the KK scale $\frac{1}{L}\rho^{-1}(L)$ plays
a role of the seesaw denominator.

If the metric factor $\rho(L)$ is used to resolve the gauge hierarchy,
the heavy-mode scales $M\rho^{-1}(L)$, $\rho^{-1}(L)/L$ are
around or smaller than TeV\@. In the present setup, there are several ways to
reproduce a proper scale of neutrino mass. A direct approach is to
consider a bulk Majorana mass of an intermediate scale. For instance,
in the case that $M\ll m_d$ and $qL\gg 1$, the neutrino mass is given 
by $M_\nu \simeq \rho(L)m^2M/m_d$ and therefore the effective
heavy-mode scale is enhanced by $m_d/M$. Thus a small lepton number
violation, $M/m_d\sim 10^{-10}$, leads to an eV-order neutrino mass.

A different type of boundary conditions for the bulk 
neutrino $\Psi(x,y)$ is a new interesting possibility for neutrino
phenomenology. Under the conditions $\Psi(x,-y)=-\gamma_5\Psi(x,y)$
and $\Psi(x,L-y)=\gamma_5\Psi(x,L+y)$, the lepton number violating
part of the correlator becomes in the low-energy regime
\begin{eqnarray}
\langle \epsilon \nu_R^*(0,y)\nu_R^\dag(0,y') \rangle
&=& \frac{M\sinh(qy_<)}
{q\Big[q\cosh(qL) + m_d \sinh(qL) \Big] } 
\nonumber\\[1mm]
&& \qquad\qquad \!\!\!
\times \Big[ q\cosh(qy_>-qL) - m_d \sinh(qy_>-qL) \Big],
\label{propZp2}
\end{eqnarray}
where the definitions of $q$ and $y_\lessgtr^{}$ are the same as 
in~\eqref{propZp}. With the propagator at hand, the neutrino 
mass reads
\begin{eqnarray}
M_\nu \;=\; \rho(L) \, 
\frac{\sinh(qL)}{ q \cosh(qL) + m_d \sinh(qL) }\;
m^{\rm T}Mm\,.
\label{mnu2}
\end{eqnarray}
This is the exact expression for taking into account of all KK-mode
contributions in the general non-factorizable geometry. Unlike the
usual seesaw mechanism, the heavy Majorana mass $M$ appears in the
numerator of~\eqref{mnu2}. In the limit $qL\gg1$, the neutrino mass
becomes $M_\nu\simeq\rho(L)\big(\frac{M}{q+m_d}\big)m^2$, which is
equivalent to $M_\nu$ for the previous boundary condition. This is
because, in the large-size limit of extra dimension $qL\gg1$, the
difference of boundary conditions at $y=0$ (for the right-handed
component) is irrelevant to the physics at another boundary where the
left-handed neutrinos reside. On the other hand, for the opposite
limit $qL\ll 1$, the neutrino mass $M_\nu$ is approximated as
$M_\nu\simeq \rho(L)LMm^2$. In the KK-mode picture,
the low-energy spectrum from $\Psi$ has no chiral zero mode and
consists of vector-like KK fermions which are perturbed by small
Majorana masses. Therefore the seesaw-induced mass from these states
is proportional to $M$, not inverse-proportional as in the usual
seesaw mechanism. Thus a small ratio of $M$ to the
compactification scale $1/L$ serves as a suppression factor for the
neutrino mass: for instance, an eV-order $M_\nu$ is
reproduced by $ML\sim 10^{-10}$.

As mentioned previously, the extension of left-handed neutrinos
(lepton doublets) to the extra dimension is also a possible way to
realize a tiny mass scale $M_\nu$ by reducing the boundary Dirac 
mass $m$ (i.e.\ the Yukawa coupling) to a smaller value than the
electroweak scale. The suppression with
the wave functions of left-handed neutrinos cannot be arbitrarily
strong since it also brings down the charged-lepton mass scale. 
To estimate to what extent their wave functions can suppress the
neutrino mass, let us assume the right-handed tau to reside on the
TeV brane. In this case, the neutrino mass $M_\nu$ is
suppressed by the factor of $(m_\tau/\text{TeV})^2\sim 10^{-6}$. 

\bigskip

%%%%%%%%%%%%%%%%%%%%%%%%%%%%%%%%%%%%%%%%%%%%%%%%%%%%%%%%%%%%%%%%%%%%%%%
\section{Summary and discussion}
%%%%%%%%%%%%%%%%%%%%%%%%%%%%%%%%%%%%%%%%%%%%%%%%%%%%%%%%%%%%%%%%%%%%%%%

We have discussed the seesaw mechanism with bulk and boundary
Majorana mass terms in the general non-factorizable geometry,
referring to the warped AdS$_5$ metric as a typical example. We have
derived the exact seesaw-induced masses by the Kaluza-Klein 
expansion and the propagator method, and presented the results in
analytic forms which make it straightforward to analyze what types of
effects are involved in the induced neutrino masses. The details of
wave functions and background geometry are irrelevant to the light
neutrino mass when the metric factor is fixed, e.g.\ by solving the
gauge hierarchy problem. The observed tiny mass scale of left-handed
neutrinos is reproduced in both setups with small lepton number
violation or other effects.

It is easy to include in the present framework the three generation
fermions and their mixture. The flavor structure is introduced with
masses and couplings in the generation space, which would be
controlled by some fundamental dynamics to be specified. Within the
higher-dimensional theory, the observed neutrino mixing is obtained,
for example, by flavor symmetry and its breaking by orbifold boundary
conditions imposed on bulk fields~\cite{TF}, where non-abelian
discrete flavor symmetry is adopted and the tri-bimaximal generation
mixing~\cite{tribi} is realized as a direct consequence of the
theory. On the other hand, it is often discussed in flavor theory that
different generations have different position profiles in the extra
dimensions~\cite{split} and, if bulk right-handed neutrinos included,
they connect up these generations across the extra dimensions. In such
cases, the propagator method presented in this letter would be useful
for finding explicit expression of light neutrino masses and collider
signatures of bulk neutrinos~\cite{collider} avoiding KK mode
sums. The unified dynamics including charged fermions and
intermediate-scale Majorana masses remains to be explored in future
work.

\bigskip
%%%%%%%%%%%%%%%%%%%%%%%%%%%%%%%%%%%%%%%%%%%%%%%%%%%%%%%%%%%%%%%%%%%%%%%
\subsection*{Acknowledgments}

This work is supported in part by the scientific grant from the
ministry of education, science, sports, and culture of Japan
(No.~20740135) and also by the grant-in-aid for the global COE program
"The next generation of physics, spun from universality and emergence"
and the grant-in-aid for the scientific research on priority area
(\#441) "Progress in elementary particle physics of the 21st century
through discoveries of Higgs boson and supersymmetry" (No.~16081209).

\bigskip\bigskip
%%%%%%%%%%%%%%%%%%%%%%%%%%%%%%%%%%%%%%%%%%%%%%%%%%%%%%%%%%%%%%%%%%%%%%%


\begin{thebibliography}{99} 

\bibitem{ADD}
%\cite{ArkaniHamed:1998rs}
%\bibitem{ArkaniHamed:1998rs}
  N.~Arkani-Hamed, S.~Dimopoulos and G.R.~Dvali,
  %``The hierarchy problem and new dimensions at a millimeter,''
  Phys.\ Lett.\ B {\bf 429} (1998) 263;
  %[arXiv:hep-ph/9803315].
  %%CITATION = PHLTA,B429,263;%%
%\cite{Antoniadis:1998ig}
%\bibitem{Antoniadis:1998ig}
 I.~Antoniadis, N.~Arkani-Hamed, S.~Dimopoulos and G.R.~Dvali,
  %``New dimensions at a millimeter to a Fermi and superstrings at a
  %TeV,''
  Phys.\ Lett.\ B {\bf 436} (1998) 257.
  %[arXiv:hep-ph/9804398].
  %%CITATION = PHLTA,B436,257;%%

\bibitem{RS}
%\cite{Randall:1999ee}
%\bibitem{Randall:1999ee}
  L.~Randall and R.~Sundrum,
  %``A large mass hierarchy from a small extra dimension,''
  Phys.\ Rev.\ Lett.\ {\bf 83} (1999) 3370;
  %[arXiv:hep-ph/9905221].
  %%CITATION = PRLTA,83,3370;%%
%\cite{Randall:1999vf}
%\bibitem{Randall:1999vf}
%  L.~Randall and R.~Sundrum,
  %``An alternative to compactification,''
  {\it ibid.} {\bf 83} (1999) 4690.
  %[arXiv:hep-th/9906064].
  %%CITATION = PRLTA,83,4690;%%

\bibitem{Psi}
%\cite{Dienes:1998sb}
  K.R.~Dienes, E.~Dudas and T.~Gherghetta,
  % ``Light neutrinos without heavy mass scales: A higher-dimensional
  % seesaw mechanism,''
  Nucl.\ Phys.\ B {\bf 557} (1999) 25;
  %[arXiv:hep-ph/9811428];
  %%CITATION = NUPHA,B557,25;%%
%\cite{ArkaniHamed:1998vp}
%\bibitem{ArkaniHamed:1998vp}
  N.~Arkani-Hamed, S.~Dimopoulos, G.R.~Dvali and J.~March-Russell,
  %``Neutrino masses from large extra dimensions,''
  Phys.\ Rev.\ D {\bf 65} (2002) 024032;
  %[arXiv:hep-ph/9811448].
  %%CITATION = PHRVA,D65,024032;%%
%\cite{Faraggi:1999bm}
%\bibitem{Faraggi:1999bm}
  A.E.~Faraggi and M.~Pospelov,
  %``Phenomenological issues in TeV scale gravity with light neutrino
  %masses,''
  Phys.\ Lett.\ B {\bf 458} (1999) 237;
  %[arXiv:hep-ph/9901299].
  %%CITATION = PHLTA,B458,237;%%
%\cite{Dvali:1999cn}
%\bibitem{ocs}
  G.R.~Dvali and A.Y.~Smirnov,
  %``Probing large extra dimensions with neutrinos,''
  Nucl.\ Phys.\ B {\bf 563} (1999) 63;
  %[arXiv:hep-ph/9904211];
  %%CITATION = NUPHA,B563,63;%%
%\cite{Mohapatra:1999zd}
%\bibitem{Mohapatra:1999zd}
  R.N.~Mohapatra, S.~Nandi and A.~Perez-Lorenzana,
  %``Neutrino masses and oscillations in models with large extra
  %dimensions,''
  Phys.\ Lett.\ B {\bf 466} (1999) 115;
  %[arXiv:hep-ph/9907520].
  %%CITATION = PHLTA,B466,115;%%
%\cite{Ioannisian:1999cw}
%\bibitem{Ioannisian:1999cw}
  A.~Ioannisian and A.~Pilaftsis,
  %``Cumulative non-decoupling effects of Kaluza-Klein neutrinos in
  %electroweak processes,''
  Phys.\ Rev.\ D {\bf 62} (2000) 066001;
  %[arXiv:hep-ph/9907522].
  %%CITATION = PHRVA,D62,066001;%%
%\cite{Barbieri:2000mg}
%\bibitem{Barbieri:2000mg}
  R.~Barbieri, P.~Creminelli and A.~Strumia,
  %``Neutrino oscillations from large extra dimensions,''
  Nucl.\ Phys.\ B {\bf 585} (2000) 28;
  %[arXiv:hep-ph/0002199].
  %%CITATION = NUPHA,B585,28;%%
%\cite{Davoudiasl:2002fq}
%\bibitem{Davoudiasl:2002fq}
  H.~Davoudiasl, P.~Langacker and M.~Perelstein,
  %``Constraints on large extra dimensions from neutrino oscillation
  %experiments,''
  Phys.\ Rev.\ D {\bf 65} (2002) 105015.
  %[arXiv:hep-ph/0201128].
  %%CITATION = PHRVA,D65,105015;%%


\bibitem{bulkF}
%\cite{Grossman:1999ra}
  Y.~Grossman and M.~Neubert,
  %``Neutrino masses and mixings in non-factorizable geometry,''
  Phys.\ Lett.\ B {\bf 474} (2000) 361;
  %[arXiv:hep-ph/9912408];
  %%CITATION = PHLTA,B474,361;%%
%\cite{Huber:2000ie}
%\bibitem{Huber:2000ie}
  S.J.~Huber and Q.~Shafi,
  %``Fermion Masses, Mixings and Proton Decay in a Randall-Sundrum Model,''
  Phys.\ Lett.\ B {\bf 498} (2001) 256;
  %[arXiv:hep-ph/0010195].
  %%CITATION = PHLTA,B498,256;%%
%\cite{Moreau:2005kz}
%\bibitem{Moreau:2005kz}
  G.~Moreau and J.I.~Silva-Marcos,
  %``Neutrinos in warped extra dimensions,''
  JHEP {\bf 0601} (2006) 048.
  %[arXiv:hep-ph/0507145].
  %%CITATION = JHEPA,0601,048;%%


\bibitem{braneM}
%\cite{Csaki:2003sh}
%\bibitem{Csaki:2003sh}
  C.~Csaki, C.~Grojean, J.~Hubisz, Y.~Shirman and J.~Terning,
  %``Fermions on an interval: Quark and lepton masses without a Higgs,''
  Phys.\ Rev.\ D {\bf 70} (2004) 015012;
  %[arXiv:hep-ph/0310355].
  %%CITATION = PHRVA,D70,015012;%%
%\cite{Perez:2008ee}
%\bibitem{Perez:2008ee}
  G.~Perez and L.~Randall,
  %``Natural Neutrino Masses and Mixings from Warped Geometry,''
  JHEP {\bf 0901} (2009) 077;
  %[arXiv:0805.4652 [hep-ph]].
  %%CITATION = JHEPA,0901,077;%%
%\cite{Carena:2009yt}
%\bibitem{Carena:2009yt}
  M.~Carena, A.D.~Medina, N.R.~Shah and C.E.M.~Wagner,
  %``Gauge-Higgs Unification, Neutrino Masses and Dark Matter in
  %Warped Extra Dimensions,''
  Phys.\ Rev.\ D {\bf 79} (2009) 096010;
  %[arXiv:0901.0609 [hep-ph]].
  %%CITATION = PHRVA,D79,096010;%%
%\cite{Chen:2009gy}
%\bibitem{Chen:2009gy}
  M.C.M.~Chen, K.T.~Mahanthappa and F.~Yu,
  %``A Viable Randall-Sundrum Model for Quarks and Leptons with T'
  %Family Symmetry,''
  arXiv:0907.3963.
  %%CITATION = ARXIV:0907.3963;%%

  
\bibitem{FN}
%\cite{Froggatt:1978nt}
%\bibitem{Froggatt:1978nt}
  C.D.~Froggatt and H.B.~Nielsen,
  %``Hierarchy Of Quark Masses, Cabibbo Angles And CP Violation,''
  Nucl.\ Phys.\ B {\bf 147} (1979) 277.
  %%CITATION = NUPHA,B147,277;%%


\bibitem{bulkM}
%\cite{Huber:2002gp}
  S.J.~Huber and Q.~Shafi,
  %``Seesaw mechanism in warped geometry,''
  Phys.\ Lett.\ B {\bf 583} (2004) 293.
  %[arXiv:hep-ph/0309252].
  %%CITATION = PHLTA,B583,293;%%


\bibitem{TF}
%\cite{Haba:2006dz}
  N.~Haba, A.~Watanabe and K.~Yoshioka,
  %``Twisted flavors and tri/bi-maximal neutrino mixing,''
  Phys.\ Rev.\ Lett.\ {\bf 97} (2006) 041601;
  %[arXiv:hep-ph/0603116];
  %%CITATION = PRLTA,97,041601;%%
% %\cite{Kobayashi:2008ih}
% \bibitem{KOY}
   T.~Kobayashi, Y.~Omura and K.~Yoshioka,
   %``Flavor Symmetry Breaking and Vacuum Alignment on Orbifolds,''
   Phys.\ Rev.\ D {\bf 78} (2008) 115006.
   %[arXiv:0809.3064 [hep-ph]].
   %%CITATION = PHRVA,D78,115006;%%


\bibitem{tribi}
%\cite{Harrison:2002er}
%\bibitem{Harrison:2002er}
  P.F.~Harrison, D.H.~Perkins and W.G.~Scott,
  %``Tri-bimaximal mixing and the neutrino oscillation data,''
  Phys.\ Lett.\ B {\bf 530} (2002) 167;
  %[arXiv:hep-ph/0202074];
  %%CITATION = PHLTA,B530,167;%%
%\cite{Harrison:2002kp}
%\bibitem{Harrison:2002kp}
  P.F.~Harrison and W.G.~Scott,
  %``Symmetries and generalisations of tri-bimaximal neutrino mixing,''
  Phys.\ Lett.\ B {\bf 535} (2002) 163.
  %[arXiv:hep-ph/0203209].
  %%CITATION = PHLTA,B535,163;%%


\bibitem{split}
%\bibitem{ArkaniHamed:1999dc}
  N.~Arkani-Hamed and M.~Schmaltz,
  %``Hierarchies without symmetries from extra dimensions,''
  Phys.\ Rev.\ D {\bf 61} (2000) 033005;
  %[arXiv:hep-ph/9903417].
  %%CITATION = PHRVA,D61,033005;%%
%\bibitem{Dvali:2000ha}
  G.R.~Dvali and M.A.~Shifman,
  %``Families as neighbors in extra dimension,''
  Phys.\ Lett.\ B {\bf 475} (2000) 295;
  %[arXiv:hep-ph/0001072].
  %%CITATION = PHLTA,B475,295;%%
%\bibitem{Yoshioka:1999ds}
  K.~Yoshioka,
  %``On fermion mass hierarchy with extra dimensions,''
  Mod.\ Phys.\ Lett.\ A {\bf 15} (2000) 29;
  %[arXiv:hep-ph/9904433].
  %%CITATION = MPLAE,A15,29;%%
%\bibitem{Hebecker:2002re}
  A.~Hebecker and J.~March-Russell,
  %``The flavour hierarchy and see-saw neutrinos from bulk masses in 5d
  %orbifold GUTs,''
  Phys.\ Lett.\ B {\bf 541} (2002) 338;
  %[arXiv:hep-ph/0205143].
  %%CITATION = PHLTA,B541,338;%%
%\bibitem{Bando:2000it}
  M.~Bando, T.~Kobayashi, T.~Noguchi and K.~Yoshioka,
  %``Yukawa hierarchy from extra dimensions and infrared fixed points,''
  Phys.\ Lett.\ B {\bf 480} (2000) 187;
  %[arXiv:hep-ph/0002102].
  %%CITATION = PHLTA,B480,187;%%
%\bibitem{Bando:2000jd}
  %M.~Bando, T.~Kobayashi, T.~Noguchi and K.~Yoshioka,
  %``Fermion mass hierarchies and small mixing angles from extra
  %dimensions,''
  Phys.\ Rev.\ D {\bf 63} (2001) 113017;
  %[arXiv:hep-ph/0008120].
  %%CITATION = PHRVA,D63,113017;%%
%\bibitem{Biggio:2003kp}
  C.~Biggio, F.~Feruglio, I.~Masina and M.~Perez-Victoria,
  %``Fermion generations, masses and mixing angles from extra
  %dimensions,''
  Nucl.\ Phys.\ B {\bf 677} (2004) 451.
  %[arXiv:hep-ph/0305129].
  %%CITATION = NUPHA,B677,451;%%


\bibitem{collider}
For example, %\cite{Cao:2004tu}
%\bibitem{Cao:2004tu}
  Q.H.~Cao, S.~Gopalakrishna and C.P.~Yuan,
  %``Collider signature of bulk neutrinos in large extra dimensions,''
  Phys.\ Rev.\ D {\bf 70} (2004) 075020;
  %[arXiv:hep-ph/0405220].
  %%CITATION = PHRVA,D70,075020;%%
%\bibitem{Haba:2009sd}
  N.~Haba, S.~Matsumoto and K.~Yoshioka,
  %``Observable Seesaw and its Collider Signatures,''
  Phys.\ Lett.\ B {\bf 677} (2009) 291;
  %[arXiv:0901.4596 [hep-ph]].
  %%CITATION = PHLTA,B677,291;%%
%\cite{Gingrich:2009az}
%\bibitem{Gingrich:2009az}
  D.M.~Gingrich,
  %``Signatures of Singlet Neutrinos in Large Extra Dimensions at the
  %LHC,''
  arXiv:0907.1878.
  %%CITATION = ARXIV:0907.1878;%%

\end{thebibliography}
\end{document}